\documentclass[aoas]{imsart}

\RequirePackage{amsthm,amsmath,amsfonts,amssymb}
\RequirePackage[authoryear]{natbib}
\RequirePackage{graphicx}
\usepackage{threeparttable}
\usepackage{multirow}
\usepackage{xcolor}
\usepackage{anyfontsize}

\startlocaldefs

\endlocaldefs

\begin{document}

\begin{frontmatter}
\title{Utilizing a Capture-Recapture Strategy to Accelerate Infectious Disease Surveillance}

\begin{aug}
\author[A,B]{\fnms{Lin}~\snm{Ge}\ead[label=e1]{lge$\_$biostat@hotmail.com}\orcid{0000-0002-8180-9615}},
\author[A]{\fnms{Yuzi}~\snm{Zhang}\ead[label=e2]{yuzi.zhang@emory.edu}\orcid{0000-0003-1805-6910}}
\author[A]{\fnms{Lance}~\snm{Waller}\ead[label=e3]{lwaller@emory.edu}\orcid{0000-0001-5002-8886}}
\and
\author[A]{\fnms{Robert}~\snm{Lyles}\ead[label=e4]{rlyles@emory.edu}}
\address[A]{Department of Biostatistics and Bioinformatics, Emory University\printead[presep={,\ }]{e2,e3,e4}}

\address[B]{Department of Biostatistics and Epidemiology, Harvard University\printead[presep={,\ }]{e1}}

\end{aug}

\begin{abstract}
Monitoring key elements of disease dynamics (e.g., prevalence, case counts) is of great importance in infectious disease prevention and control, as emphasized during the COVID-19 pandemic. To facilitate this effort, we propose a new capture-recapture (CRC) analysis strategy that adjusts for misclassification stemming from the use of easily administered but imperfect diagnostic test kits, such as Rapid Antigen Test-kits or saliva tests. Our method is based on a recently proposed “anchor stream” design, whereby an existing voluntary surveillance data stream is augmented by a smaller and judiciously drawn random sample. It incorporates manufacturer-specified sensitivity and specificity parameters to account for imperfect diagnostic results in one or both data streams. For inference to accompany case count estimation, we improve upon traditional Wald-type confidence intervals by developing an adapted Bayesian credible interval for the CRC estimator that yields favorable frequentist coverage properties. When feasible, the proposed design and analytic strategy provides a more efficient solution than traditional CRC methods or random sampling-based bias-corrected estimation to monitor disease prevalence while accounting for misclassification. We demonstrate the benefits of this approach through simulation studies and a numerical example that underscore its potential utility in practice for economical disease monitoring among a registered closed population.
\end{abstract}

\begin{keyword}
\kwd{Credible Interval}
\kwd{Misclassification}
\kwd{Non-representative Sampling}
\kwd{Sensitivity}
\kwd{Specificity}
\end{keyword}

\end{frontmatter}


\section{Introduction}

Spurred by the COVID-19 pandemic, healthcare experts, policy makers, and government administrators have become increasingly aware of the importance of infectious disease monitoring in a particular geographic region \citep{Weinberg2003,Hopkins2005}, densely populated district \citep{Soh2012b}, or vulnerable community \citep{Suyama2003}. Applying regular disease surveillance efforts among such populations can help assess the prevalence and alert policy makers of the need to address an emerging or worsening crisis. However, many voluntary-based epidemiological surveillance programs produce biased data, as they often oversample positive cases \citep{Menni2020}. A common example arose during voluntary testing programs on university campuses during the recent pandemic \citep{Rennert2021,Schultes2021,Rennert2022,VanderSchaaf2021,Matheson2021}, as students, staff and faculty were more likely to seek testing if they were feeling sick or had recent contact with active cases. That is, people with symptoms or health concerns may be more likely to participate in passive surveillance surveys, leading to overestimation of true disease prevalence in a closed community.  

In epidemiology or public health-related surveillance research, the capture-recapture (CRC) approach, which was borrowed from ecology studies to estimate the size of wildlife populations, is now commonly advocated for estimating case counts and prevalences. Applications of CRC have been directed toward many infectious diseases, such as HIV \citep{Poorolajal2017}, Hepatitis C \citep{Wu2005} and tuberculosis \citep{Dunbar2011,Carvalho2020,PerezDuque2020}. For accurate estimation, one key assumption that is often made is that there are no population-level associations among the surveillance efforts used; this is known as the Lincoln-Petersen, or “LP” condition in two-stream CRC analysis. Classic tools such as the Lincoln-Petersen \citep{Lincoln1930,Petersen1986} and Chapman estimators \citep{Chapman1951} are built on this assumption. However, it is often questionable in practice, and violating it may lead to biased estimation of the prevalence or population size \citep{Brenner1995}. While great effort has been directed toward relaxation of such assumptions, many sources \citep{Agresti1994,Hook1995,Cormack1999} point out that applying popular CRC estimation strategies in practice is almost always fraught with pitfalls; this includes significant drawbacks to the popular loglinear modeling paradigm \citep{Fienberg1972,Baillargeon2007,Jones2014,Zhang2023b}. To better explore relationships between multiple CRC data sources, some researchers \citep{Chatterjee2016,Zhang2020,Zhang2023} have proposed sensitivity analysis to evaluate the uncertainty stemming from unknown levels of association among surveillance streams. However, it is generally recognized that a design-based approach would be the only sure-fire way to ensure the LP condition in practice \citep{Seber1982,Chao2008,Lyles2022a, Lyles2022b}. When feasible, this approach achieves the crucial requirement by introducing a second random sampling-based surveillance effort that is implemented carefully so as to provide case identification independently of the existing non-representative disease surveillance data stream \citep{Lyles2022a}. When it can be appropriately implemented in a closed and enumerable population, this sampling strategy leads to an unbiased maximum likelihood (ML) estimator of the case count that is typically far more precise than classical CRC estimators derived under the LP condition. This comes about on the strength of a so-called “anchor stream” design, which precisely identifies a crucial conditional sampling probability parameter associated with the random sampling-based data stream \citep{Lyles2022a,Lyles2022b}.

A common challenge when analyzing epidemiological surveillance data is that the diagnostic method for ascertaining disease status may be prone to error. That is, the diagnostic results observed in disease surveillance programs may rely on imperfect tests or diagnostic devices,  which can lead to misclassification errors \citep{Rennert2021,Schultes2021,Rennert2022,VanderSchaaf2021,Matheson2021}. Although an imperfect test result can lead to biased estimation, it is often the case that no gold standard is available to assess presence or absence of a particular disease \citep{Glasziou2008,Walter2012}. On the other hand, even when an accurate diagnostic test exists, some common but imperfect tests offer benefits such as ease of application, immediacy of results, and the ability to quickly assess quality improvement plans during the epidemiological disease screening process. However, these tests will indeed generally suffer from a lack of gold-standard accuracy and sensitivity \citep{Soh2012}. Regarding the fallible disease status indications obtained from a single random sampling-based data source, numerous studies \citep{Levy1970,Rogan1978,Gastwirth1987} have offered feasible solutions by incorporating known or estimated misclassification parameters, such as the sensitivity ($Se$) and specificity ($Sp$). Yet, few researchers have discussed this issue under the CRC paradigm, particularly when dealing with disease surveillance data. When assuming the false-positive and false-negative rates are known, \cite{Brenner1996} and \cite{Ramos2020} developed methods to adjust the error-prone surveillance data streams. \cite{Ge2023} recently proposed a generalized anchor stream design to account for misclassification errors in an existing non-representative surveillance stream, incorporating the CRC paradigm to identify an estimable positive predictive value ($PPV$) parameter to facilitate estimating the cumulative incidence of breast cancer recurrence among a select population subsetted from the Georgia Cancer Registry-based Cancer Recurrence Information and Surveillance Program (CRISP).  

In this article, we propose a CRC strategy to leverage an existing general disease surveillance effort, supplemented by what can be a relatively small random sample. Our approach is based on an extension and generalization of previously proposed methods rooted in the anchor stream design \citep{Lyles2022a,Lyles2022b,Ge2023}, in order to target unbiased estimation of disease prevalence while accounting for imperfect disease diagnoses based, for example, on Rapid Antigen test-kits or Saliva-based tests that have been commonly applied during the COVID-19 pandemic \citep{Rennert2021,Schultes2021,Rennert2022,VanderSchaaf2021,Matheson2021}. This strategy allows for the estimation of disease case counts within a closed population region or community. It justifies fallible diagnostic status indications obtained via both data streams by leveraging manufacturer-reported sensitivity and specificity information, while preserving the independence and random sampling properties of the anchor stream. Importantly, we thus relax the strict stipulation requiring accurate test results in prior proposals of the anchor stream design \citep{Lyles2022a,Lyles2022b} in such disease monitoring settings, to accommodate imperfect diagnostic results via both data streams. In turn, this extension allows for the potential acceleration of epidemiological surveillance programs during an infectious disease season or pandemic. 

\section{Methods}\label{section_2}

\subsection{Misclassification Parameters}

The two misclassification parameters $Se$ and $Sp$ are integral to developing and assessing diagnostic tests, as they quantify the proportion of the test’s positive and negative results that are true positives and true negatives. Sensitivity is the probability of a positive test result given the tested individual is truly diseased, i.e., $Se$ = Pr(Test positive $|$ diseased). Specificity is the probability of a negative test result given the tested individual is truly non-diseased, i.e., $Sp$ = Pr(Test negative $|$ non-diseased). Conversely, false positive results are defined as Pr(Test positive $|$ non-diseased) = 1-$Sp$, and false negative results as Pr(Test negative $|$ diseased) = 1-$Se$. In what follows, we use these familiar parameters to adjust for misclassified disease status.

\subsection{Anchor Stream Design}\label{section_2.2}

{This study is motivated by the need to conduct surveillance of infectious diseases during an outbreak, in order to estimate the current case count or prevalence within a sizable yet essentially closed population. A relevant example is the monitoring of COVID-19 across university campuses during the recent pandemic \citep{Rennert2021,Schultes2021,Rennert2022,VanderSchaaf2021,Matheson2021}. A first data source for this surveillance would be derived from voluntary testing programs implemented on campus. Recent articles promoting an “anchor stream” design \citep{Lyles2022a,Lyles2022b} suggest augmenting such voluntary programs by randomly selecting and testing a comparatively small sample of individuals at the close of the monitoring period. With an available list and access to target population members, \cite{Lyles2022a} discuss how this design can be leveraged toward estimating cumulative incidence of new cases over a defined time period, or toward estimation of current (e.g., as of a given day) disease prevalence. The key to ensuring the necessary independence between the existing non-representative data stream and the design-based anchor stream is to implement the latter in a “post-enumeration” fashion, so that notification of random selection does not impact one’s choice about whether to seek a voluntary test.} 

We build on prior considerations of the anchor stream design without misclassification \citep{Lyles2022a,Lyles2022b} along with extensions that proposed a justified CRC estimator based on use of an error-free disease assessment to access an estimable PPV for targeting cumulatively incident case counts \citep{Ge2023}. Here, we focus instead on the goal of estimating the current case count to surveil disease within an enumerated registry population without the stipulation that the anchor stream must employ a perfect diagnostic testing method. The existing surveillance effort, referred to as Stream 1, typically selects those at high risk of disease preferentially and is also likely to use an error-prone testing method. We subsequently obtain a random sample of individuals from the registered target population as the “anchor stream,” or Stream 2, which (as noted previously) is carefully designed to be agnostic (independent) of Stream 1 \citep{Lyles2022a,Lyles2022b}. Importantly, we allow each of the two surveillance efforts to be based on its own error-prone diagnostic method characterized by known $Se$ and $Sp$ values provided by the manufacturer of the diagnostic device or test-kit.  

Benefiting from this design, the anchor stream alone provides its own valid and defensible sampling-based estimator based on known manufacturer-specified  $Se$ and $Sp$ \citep{Ge2023b}. However, Stream 2 is typically expected to include a relatively small sample size, and is likely to identify far fewer potential cases relative to Stream 1. Assuming the total population size ($N_{tot}$) of the closed community or registry is known in advance, the bias-corrected true prevalence estimator  $\pi_c$ and the corresponding case count estimator ($\hat{N}_{RS}$) based on the Stream 2 random sample with size ($n$) and known sensitivity ($Se_2$) and specificity ($Sp_2$) are given by the following formulae \citep{Rogan1978,Gastwirth1987,Levy1970,Ge2023b}: 
\begin{align}
    \hat{\pi}_c = \frac{\hat{\pi}+Sp_2-1}{Se_2+Sp_2-1}, ~~ \hat{N}_{RS}=N_{tot}\hat{\pi}_c, ~~ \hat{V}(\hat{N}_{RS}) = N_{tot}^2\hat{V}(\hat{\pi}_c) \label{eq_3.1}
\end{align}
where $\hat{\pi}=n^+/n$, and $n^+$ denotes the number of individuals identified as test positives in the random sample. When calculating the bias-corrected prevalence estimator $\hat{\pi}_c$, one needs to consider a threshold justification as follows \citep{Ge2023b} in light of the natural constraint $1-Sp_2\leq \hat{\pi}\leq Se_2$ that exists in the general error-prone testing problem:
\begin{align}\label{eq_2}
    \hat{\pi}_c = 
     \begin{cases}
      0 & \hat{\pi}\leq 1-Sp_2 \\
      1 & \hat{\pi}\geq Se_2   \\
      \hat{\pi}_c & \text{else}
    \end{cases}     
\end{align}

Given that the total population is closed and finite, a recently developed variance estimator $\hat{V}(\hat{\pi}_c)$ \citep{Ge2023b} incorporates a finite population correction (FPC) given by Cochran \citep{Cochran1977} together with an elusive but necessary second term, i.e.,
\begin{align}\label{eq_3}
    \hat{V}(\hat{\pi}_c) =& \frac{1}{(Se_2+Sp_2-1)^2}\big\{ \big[\frac{n(N_{tot}-n)}{N_{tot}(n-1)} \big] \frac{\hat{\pi}(1-\hat{\pi})}{n} + \nonumber\\
    & \frac{1}{N_{tot}}[\hat{\pi}_c Se_2(1-Se_2)+(1-\hat{\pi}_c)Sp_2(1-Sp_2)]\big\}
\end{align}

When the anchor stream applies a perfect test (i.e., $Se_2=Sp_2=1$), the variance estimator in equation (\ref{eq_3}) reduces to the standard FPC-corrected sampling-based variance estimator, i.e., $\hat{V}(\hat{\pi}_c) = \big[\frac{n(N_{tot}-n)}{N_{tot}(n-1)} \big] \frac{\hat{\pi}(1-\hat{\pi})}{n}$. Moreover, when the total population size $N_{tot}$ is relatively small and the anchor stream sample size $n$ is large in comparison to $N_{tot}$, the finite population effect leads to a substantial reduction in variance.

\subsection{A Novel Capture-Recapture (CRC) Estimator}\label{section_2.3}

We now assume that the disease assessment methods applied via the anchor stream design are fallible in both data streams, with known Sensitivity ($Se_1$, $Se_2$) and Specificity ($Sp_1$, $Sp_2$). A novel CRC estimator using all available data is justified using maximum likelihood (ML) based on a general multinomial model for the nine cell counts defined in Table \ref{table_1}.

\begin{table}
\centering
\caption{Cell Counts and Likelihood Contributions}
\label{table_1}
 \begin{tabular}{cll} 
 \hline
  Cell & \multicolumn{1}{c}{Observation Type} & \multicolumn{1}{c}{Likelihood Contributions} \\ 
 \hline
 $n_1$ & Sampled in Both Streams, Test + in Both   & $p_1 = \psi[Se_2 Se_1 \pi_{1} + (1-Sp_2)(1-Sp_1) (1-\pi_{1})]\phi$\\ 
 $n_2$ & Sampled in Both Streams, Test $-$ in Both & $p_2 = \psi[(1-Se_2) (1-Se_1)\pi_{1} + Sp_2 Sp_1 (1-\pi_{1})]\phi$\\
 \multirow{2}{*}{$n_3$} & Sampled in Both Streams, Test + in Stream 1,& \multirow{2}{*}{$p_3 = \psi[(1-Se_2) Se_1\pi_{1} + Sp_2 (1-Sp_1)(1-\pi_{1})]\phi$}\\
  &  Test $-$ in Stream 2  &\\ 
 \multirow{2}{*}{$n_4$} & Sampled in Both Streams, Test $-$ in Stream 1, & \multirow{2}{*}{$p_4 = \psi[Se_2 (1-Se_1)\pi_{1} + (1-Sp_2) Sp_1(1-\pi_{1})]\phi$}\\
  & Test + in Stream 2  \\
 $n_5$ & Sampled in Stream 1, not Stream 2, Test +  & $p_5 = (1-\psi) [Se_1\pi_{1} + (1-Sp_1)(1-\pi_{1})]\phi$\\ 
 $n_6$ & Sampled in Stream 1, not Stream 2, Test $-$  & $p_6 = (1-\psi) [(1-Se_1)\pi_{1} + Sp_1(1-\pi_{1})]\phi$\\
 $n_7$ & Sampled in Stream 2, not Stream 1, Test +  & $p_7 =  \psi[Se_2\pi_{01} + (1-Sp_2)(1-\pi_{01})](1-\phi)$\\ 
 $n_8$ & Sampled in Stream 2, not Stream 1, Test $-$  & $p_8 =  \psi[(1-Se_2)\pi_{01} + Sp_2 (1-\pi_{01})](1-\phi)$\\
 $n_9$ & Not Sampled in Stream 1 or Stream 2  &
 $p_9 = (1-\psi)(1-\phi)$\\
 \hline
 \end{tabular}
\end{table}

The likelihood contributions presented in Table \ref{table_1} are based on defining the parameters, $\phi=$Pr(sampled in Stream 1), $\pi_{1}=$Pr(true + $|$ sampled in Stream 1), $\pi_{01}=$Pr(true + $|$ not sampled in Stream 1). In addition, we have a known parameter $\psi=$ Pr(sampled in Stream 2), which is under the investigator’s control and can be fixed as the proportion of the $N_{tot}$ individuals represented in Stream 2. While it is assumed at this point that the sensitivity and specificity parameters are known, the subscripts reflect the fact that both can differ across surveillance efforts (i.e., different testing methods can be applied in Stream 1 and Stream 2). When both disease assessments are accurate, meaning that all 4 $Se$/$Sp$ parameters can be assumed equal to 1, cell counts $n_3$ and $n_4$ and their likelihood contributions in Table \ref{table_1} will be zero. In that case, the estimators previously proposed by \cite{Lyles2022b} can be applied directly for case count estimation.

For the purpose of point estimation of the true prevalence or case count, the vector of nine cell counts in Table \ref{table_1} can be modeled as a multinomial sample with likelihood proportional to $\prod_{j=1}^9 p_j^{n_j}$, where $p_j$ denotes the likelihood contribution corresponding to the $j$th cell. That is, for point estimation one can assume
\begin{align}
    (n_{1},n_{2}, \cdots, n_{9})\sim~multinominal(N_{tot}; p_{1},p_{2}, \cdots, p_{9})
\end{align}

The MLE for the unknown parameters in Table \ref{table_1} can be obtained numerically, and we find that two of them are available in closed form. {The exception is the parameter $\pi_1$, for which one can obtain a numerical approximation ($\hat{\pi}_1^*$) by maximizing the likelihood via popular optimization packages (e.g., ``\textit{optim}'' in R). The MLEs for the other parameters in Table \ref{table_1} are in closed forms, as follows.}
\begin{align*}
    &\hat{\phi} = \frac{n_1+n_2+n_3+n_4+n_5+n_6}{N_{tot}} \\
    &\hat{\pi}_{01} = \frac{\frac{n_7}{n_7+n_8}+Sp_2-1}{Se_2+Sp_2-1}
\end{align*}

{The overall disease prevalence is a function of these parameters, and thus an initial CRC estimator for the disease case count is derived accordingly:}
\begin{align}
    \hat{N}_{CRC}^* = N_{tot}[\hat{\pi}_{1}^*\hat{\phi} + \hat{\pi}_{01}(1 - \hat{\phi})].
    \label{eq_5}
\end{align}

Importantly, however, the variance-covariance matrix implied by a multinomial model for the cell counts in Table \ref{table_1} ignores standard and non-standard FPC effects that are in play under the anchor stream design. For this reason, a traditional multivariate delta method approach applied to the estimator in equation (\ref{eq_5}) while assuming the multinomial covariance structure will tend to overestimate the variance unless both data streams sample only a small proportion of the $N_{tot}$ individuals in the finite target population. {Therefore, while we examine the numerically facilitated CRC point estimator in equation (\ref{eq_5}) in subsequent empirical studies, we do not directly pursue variance estimation in conjunction with it. Instead, an approximation approach is introduced below to facilitate a closed-form point estimate and its variance. }


In order to accommodate FPC adjustments, we first tailor the estimator of $\pi_1$ by approximating it via $\psi\hat{\pi}_{11}+(1-\psi)\hat{\pi}_{10}$, where $\hat{\pi}_{11}= \frac{\frac{n_1+n_4}{n_1+n_2+n_3+n_4}+Sp_2-1}{Se_2+Sp_2-1}$ and $\hat{\pi}_{10}=\frac{\frac{n_5}{n_5+n_6}+Sp_1-1}{Se_1+Sp_1-1}$ are estimates of the prevalence among individuals sampled by both data streams, and individuals only sampled by Stream 1, respectively. This leads to a second closed-form estimator, which compares well empirically with equation (\ref{eq_5}) across a broad range of conditions:
\begin{align}
     \hat{N}_{CRC} = N_{tot}[\psi\hat{\pi}_{11}\hat{\phi}+(1-\psi)\hat{\pi}_{10}\hat{\phi} + \hat{\pi}_{01}(1 - \hat{\phi})]
    \label{eq_6}   
\end{align}

We subsequently make use of two variance approximations for the CRC estimator in (\ref{eq_6}), as follows:
\begin{align}
    \hat{V}_{k}(\hat{N}_{CRC}) = N_{tot}^2[\hat{d}_{11}^2\hat{V}_k(\hat{\pi}_{11})+\hat{d}_{10}^2\hat{V}_k(\hat{\pi}_{10})+&\hat{d}_{01}^2\hat{V}_k(\hat{\pi}_{01})], k = 1,2\label{eq_7}
\end{align}
where $\hat{d}_{11}=\psi\hat{\phi}$, $\hat{d}_{10}=(1-\psi)\hat{\phi}$, {$\hat{d}_{01}=1-\hat{\phi}$}. For $k=1$, the approximate variance incorporates no FPC adjustments, i.e., $\hat{V}_1(\hat{\pi}_{11}) = \frac{1}{(Se_2+Sp_2-1)^2}\frac{\Tilde{\pi}_{11}(1-\Tilde{\pi}_{11})}{n_1+n_2+n_3+n_4}$, $\Tilde{\pi}_{11} = \frac{n_1+n_4}{n_1+n_2+n_3+n_4}$; $\hat{V}_1(\hat{\pi}_{10})=\frac{1}{(Se_1+Sp_1-1)^2}\frac{\Tilde{\pi}_{10}(1-\Tilde{\pi}_{10})}{n_5+n_6}$, $\Tilde{\pi}_{10}=\frac{n_5}{n_5+n_6}$; $\hat{V}_1(\hat{\pi}_{01})=\frac{1}{(Se_2+Sp_2-1)^2}\frac{\Tilde{\pi}_{01}(1-\Tilde{\pi}_{01})}{n_7+n_8}$, $\Tilde{\pi}_{01}=\frac{n_7}{n_7+n_8}$.

Thus the variance estimator $\hat{V}_{1}(\hat{N}_{CRC})$ is a conservative approximation for the variance of (\ref{eq_6}) based on a tailored version of the multivariate delta method that assumes a standard multinomial covariance structure applies to Table \ref{table_1}. In contrast, the scenario where $k=2$ incorporates FPC adjustments \citep{Cochran1977} together with the misclassification effect adjustments in (\ref{eq_3}), applying them to $\hat{V}_1(\hat{\pi}_{11})$, $\hat{V}_1(\hat{\pi}_{10})$ and $\hat{V}_1(\hat{\pi}_{01})$ in (\ref{eq_7}). That is,
\begin{align}
    \hat{V}_2(\hat{\pi}_{ij}) = FPC_{ij}\hat{V}_1(\hat{\pi}_{ij}) + \hat{V}_{extra}^{ij}, ~~ i,j=0,1 \label{eq_8}
\end{align}
where $FPC_{11}=\frac{N_{11}(N_1-N_{11})}{N_1(N_{11}-1)}$, $FPC_{10}=\frac{(N_1-N_{11})N_{11}}{N_1(N_1-N_{11}-1)}$, $FPC_{01}=\frac{N_{01}(N_{tot}-N_1-N_{01})}{(N_{tot}-N_1)(N_{01}-1)}$, $N_1=n_1+n_2+n_3+n_4+n_5+n_6$, $N_{11}=n_1+n_2+n_3+n_4$, $N_{01}=n_7+n_8$. The details of the extra variance terms ($\hat{V}_{extra}^{ij}$) are available in Appendix A of the Supplementary Material \citep{Ge2024}. This provides an alternative FPC-adjusted variance estimator, $\hat{V}_{2}(\hat{N}_{CRC})$, which we recommend for use in conjunction with the CRC estimator in (\ref{eq_6}).

{When calculating the prevalence estimators $\hat{\pi}_{ij}$ for use in the variance calculations outlined above, we apply the natural [0, 1] constraint to each of them. For the closed-form CRC estimator in (6), we recommend thresholding the overall prevalence estimate (within the square brackets) to fall within the [0, 1] interval in rare instances in which it is negative or exceeds one. }

\subsection{An Adapted Bayesian Credible Interval Approach for Inference}\label{section_2.4}

Many references have pointed out that Wald-type confidence intervals (CIs) often show poor performance when proportions are extreme and/or the sample size is limited \citep{Ghosh1979,Blyth1983,Agresti1998,Brown2001}. To potentially improve the frequentist coverage properties of the intervals accompanying the CRC estimator (\ref{eq_6}) for disease case counts {(especially in limited sample size scenarios)} while adjusting the variance for finite population effects, we adopt a Bayesian credible interval based on a weakly informative Dirichlet prior on a multinomial model.

Our approach is similar in spirit to a recent proposal for the case of no misclassification \citep{Lyles2022b}. Specifically, we implement a scale and shift adjustment to a typical posterior credible interval for $\hat{N}_{CRC}$ based on a Jeffreys' Dirichlet(1/2,~1/2,~$\cdots$,1/2) prior for the cell probabilities in Table \ref{table_1}, which yields the corresponding posterior distribution in (\ref{eq_3.16}):
\begin{align}
    Dirichlet(n_1+\frac{1}{2}, n_2+\frac{1}{2}, \cdots, n_9+\frac{1}{2}) \label{eq_3.16}
\end{align}

The traditional 95\% credible interval is defined using 2.5th and 97.5th percentiles of the target estimand in (\ref{eq_6}) based on this posterior distribution via posterior samples, i.e., $\hat{N}_{CRC}^{(s)}$, $s=1,2,\cdots,S$. To adjust the variance for finite population effects, we define a new \textit{scale} parameter $a$ and a \textit{shift} parameter $b$ as follows:
\begin{align}
    a^{(s)} = \sqrt{\hat{V}_{2}(\hat{N}_{CRC}^{(s)})/\hat{V}_{1}(\hat{N}_{CRC}^{(s)})}, ~~b^{(s)}=\hat{N}_{CRC}(1-a^{(s)}) \label{eq_3.17}
\end{align}
where $\hat{V}_{1}(\hat{N}_{CRC}^{(s)})$ and $\hat{V}_{2}(\hat{N}_{CRC}^{(s)})$ are the estimated unadjusted variance and FPC-adjusted variance for $\hat{N}_{CRC}^{(s)}$ based on applying equation (\ref{eq_7}) to the $s$-th set of posterior-sampled cell counts. Posterior draws $\hat{N}_{CRC}^{(s)}$ are then scaled and shifted, i.e.,
\begin{align}
    \Tilde{N}_{CRC}^{(s)} = a^{(s)}\hat{N}_{CRC}^{(s)} + b^{(s)} \label{eq_11}
\end{align}

This adjusts the posterior distribution to have a mean equal to $\hat{N}_{CRC}$ and incorporates adjustments to the variance for finite population and misclassification effects. We refer to the interval ($LL_{ab}, LL_{ab}$) as the proposed Bayesian credible interval for $\hat{N}_{CRC}$ by taking the 2.5\% and 97.5\% percentiles from the posterior draws in (\ref{eq_11}).

While the proposed Bayesian credible interval will typically be narrower than alternatives based on Stream 2 only, it can be conservative under certain conditions (e.g., if the Stream 2 sampling rate is large). As a comparator, we recommend examining the Bayesian credible interval proposed by \cite{Ge2023} for accompanying the Stream 2 only random sampling-based estimator $\hat{\pi}_c$ in (\ref{eq_3.1}) under finite sampling conditions; we refer to the corresponding interval for the case count as $N_{tot}\times (LL_{RS}, UL_{RS})$. In practice, we promote the use of the narrower of this interval and the interval based on equation (\ref{eq_11}); this approach is evaluated in our subsequent simulation studies. 


\subsection{{A Multiple Imputation (MI)-based Approach for Estimable $Se$ and $Sp$ parameters}}\label{section_2.5}

While the methods introduced above are based upon the assumption of known $Se$ and $Sp$ values, this will not always be realistic in practice. Hence, we introduce a multiple imputation (MI)-based approach \citep{Rubin1987} to relax this assumption in the event that external validation data are available to facilitate the estimation of these parameters. In Section \ref{section_4}, a real external validation data set is presented in Table \ref{table_5} as an example to illustrate this practical scenario.

To elaborate this approach, we firstly treat the $Se$ and $Sp$ parameters as missing and impute them using the Dirichlet distribution as follows:
\begin{align}
    (v_{11,g}^*,v_{10,g}^*,v_{01,g}^*,v_{00,g}^*) \sim     Dirichlet(v_{11,g}+\frac{1}{2},v_{10,g}+\frac{1}{2},v_{01,g}+\frac{1}{2},v_{00,g}+\frac{1}{2})
\end{align}
where $(v_{11,g},v_{10,g},v_{01,g},v_{00,g})$ are cell counts comprising the validation data for test $g$ as exemplified in Table \ref{table_5}, and $g=1,2$ indicates the data for Stream 1 or Stream 2. Corresponding draws of the sensitivity and specificity parameters are obtained as $Se_g^*=\frac{v_{11,g}^*}{v_{11,g}^*+v_{10,g}^*}$ and $Sp_g^*=\frac{v_{00,g}^*}{v_{01,g}^*+v_{00,g}^*}$. Suppose we have $M>1$ independent imputations for each of the four $Se$ and $Sp$ parameters, thereby obtaining $M$ estimates $\hat{N}_{CRC}^{(m)}$ and $\hat{N}_{RS}^{(m)}$, $m=1,2,\cdots,M$. The corresponding within-imputation variance estimates $\hat{U}_{CRC}^{(m)}$ and $\hat{U}_{RS}^{(m)}$, $m=1,2,\cdots,M$, are evaluated via the original estimators introduced in Section \ref{section_2.2} and \ref{section_2.3} and the multiple imputation estimators are calculated by taking the average of all imputations, i.e., $\hat{N}_{CRC, MI} = \frac{1}{M}\sum\hat{N}_{CRC}^{(m)}$ and $\hat{N}_{RS, MI} = \frac{1}{M}\sum\hat{N}_{RS}^{(m)}$. The MI-based variance is then calculated in standard fashion \citep{Rubin1987} as $\hat{V}_{MI}=(1+M^{-1})B+\Bar{U}$, where $B$ is the sample variance across imputations and $\Bar{U}$ is the sample mean of the within-imputation variances.

As a corresponding extension of the proposed Bayesian credible interval approach, we generate posterior samples of $\Tilde{N}_{CRC}^{(s_m)}$, $s_m=1,2,\cdots,S$; $m=1,2,\cdots,M$ based on equation (\ref{eq_11}) in Section \ref{section_2.4} along with each imputation set of $Se$ and $Sp$ parameters under the MI paradigm. We then calculate the Bayesian credible interval to accompany $\hat{N}_{CRC,MI}$ by pooling all posterior samples together.

\section{Simulation Study}\label{section_3}

We conducted simulations to assess the properties of the proposed case count estimators of $N$ along with the proposed credible interval approach, across a wide range of parameter settings. The population size $N_{tot}$ was set to 200, {1,000 and 10,000}, while the true disease prevalence was varied over a range ($p$=0.1, 0.3, 0.5). Data were generated in such a way that among those with disease, 50\% of individuals exhibited symptoms. In contrast, only 10\% of those without disease showed symptoms. The Stream 1 sample was drawn to reflect voluntary-based non-representative surveillance data, selecting 80\% of individuals with symptoms for testing as opposed to 10\% of those without symptoms. Stream 2 was generated as the anchor stream independently of Stream 1, with the {sample size} varied over a wide range {($n_2$ = 10, 20, 60) for $N_{tot}=$ 200 and ($n_2$ = 50, 100, 300) for $N_{tot}=$ 1,000 and 10,000}. Both streams included misclassified diagnostic results, controlled by known parameters ($Se_1$, $Sp_1$) and ($Se_2$, $Sp_2$) to produce a range of “low”, “medium” and “high” levels of misclassification (e.g., $Se$, $Sp$=0.95, 0.9, 0.85). We conducted 5,000 simulations for each setting, and we report results for the proposed Bayesian credible intervals for inference based on 1,000 Dirichlet posterior draws.   

\begin{table}
    \centering
    \caption{Comparing the Performance of Estimators with $N_{tot}$=1,000 and Low Misclassification Level $^a$ }
    \label{table_2}
    \begin{tabular*}{5in}{@{\extracolsep{\fill}}cccccccc@{\extracolsep{\fill}}}
    \hline
    Prevalence  & Sample & \multirow{2}{*}{Estimator $^{b}$} & \multirow{2}{*}{Mean} & \multirow{2}{*}{SD} & \multirow{2}{*}{Avg. SE} & \multirow{2}{*}{Avg. width $^{c}$} & CI Coverage\\
    $p$ & Size $(n_2)$ & & & & & &  (\%)\\
\hline
    
    & \multirow{3}{*}{50} & $\hat{N}_{RS}$ & 101.9 & 53.8 & 53.3 & 209.0         & 92.9  \\ 
    & ~ & $\hat{N}_{CRC}$ & 100.9 & 44.6 & 43.5 & 170.4 \textbf{(169.7)} & 92.4 \textbf{(93.8)}  \\ 
    & ~ & $\hat{N}_{CRC}^*$ & 102.3 & 42.5 & - & - & -   \\ \cline{2-8}

    & \multirow{3}{*}{100}  & $\hat{N}_{RS}$ & 100.2 & 38.2 & 37.2 & 145.8 & 93.3   \\ 
    0.1 & ~ & $\hat{N}_{CRC}$ & 100.3 & 31.0 & 30.8 & 120.7 \textbf{(116.6)} & 92.8 \textbf{(94.1)}   \\ 
    ~ & ~ & $\hat{N}_{CRC}^*$ & 100.3 & 31.0 & - & - & -   \\ \cline{2-8}
    
     & \multirow{3}{*}{300}  & $\hat{N}_{RS}$ & 100.4 & 20.1 & 20.1 & 79.0 & 94.3  \\ 
     & ~ & $\hat{N}_{CRC}$ & 100.3 & 17.1 & 17.7 & 69.5 \textbf{(69.5)} & 95.4 \textbf{(95.6)}  \\ 
    & ~ & $\hat{N}_{CRC}^*$ & 100.3 & 17.0 & - & - & -   \\  \cline{1-8}

    & \multirow{3}{*}{50}  & $\hat{N}_{RS}$ & 299.3 & 72.1 & 71.7 & 281.1 & 93.4   \\ 
     & ~ & $\hat{N}_{CRC}$ & 300.1 & 58.7 & 57.8 & 226.4 \textbf{(220.9)} & 92.8 \textbf{(94.5)}   \\ 
    ~ & ~ & $\hat{N}_{CRC}^*$ &  300.1 & 58.7  & - & - & -   \\ \cline{2-8}
    
    & \multirow{3}{*}{100} & $\hat{N}_{RS}$ & 298.4 & 49.6 & 49.7 & 194.8 & 95.3  \\ 
     0.3 & ~ & $\hat{N}_{CRC}$ &  298.5 & 39.8 & 40.4 & 158.2 \textbf{(156.6)} & 94.4 \textbf{(95.0)}  \\ 
    ~ & ~ & $\hat{N}_{CRC}^*$ & 298.5 & 39.8 & - & - & -   \\ \cline{2-8}
    
     & \multirow{3}{*}{300}  & $\hat{N}_{RS}$ & 300.6 & 26.4 & 26.2 & 102.7 & 95.2  \\ 
     & ~ & $\hat{N}_{CRC}$ & 300.1 & 21.6 & 22.4 & 87.9 \textbf{(88.3)} & 95.6 \textbf{(95.7)}  \\ 
      & ~ & $\hat{N}_{CRC}^*$ & 300.2 & 21.6 & - & - & -   \\ \cline{1-8}

    & \multirow{3}{*}{50}  & $\hat{N}_{RS}$ & 500.5 & 76.3 & 77.0 & 301.8 & 94.4   \\ 
     & ~ & $\hat{N}_{CRC}$ & 500.0 & 63.2 & 63.5 & 248.7 \textbf{(240.4)} & 94.3 \textbf{(95.1)}   \\ 
    ~ & ~ & $\hat{N}_{CRC}^*$ &  500.0 & 16.4   & - & - & -   \\ \cline{2-8}
    
    & \multirow{3}{*}{100} & $\hat{N}_{RS}$ & 499.3 & 53.6 & 53.3 & 208.8 & 95.3  \\ 
    0.5   & ~ & $\hat{N}_{CRC}$ &  499.3 & 44.5 & 44.1 & 173.0 \textbf{(170.3)} & 94.5 \textbf{(95.0)}  \\ 
    ~ & ~ & $\hat{N}_{CRC}^*$ & 499.4 & 44.5 & - & - & -   \\ \cline{2-8}
    
    & \multirow{3}{*}{300}  & $\hat{N}_{RS}$ & 500.6 & 27.9 & 27.9 & 109.4 & 94.5  \\ 
    & ~ & $\hat{N}_{CRC}$ & 500.3 & 23.4 & 24.2 & 94.7 \textbf{(95.1)} & 95.8 \textbf{(96.0)} \\ 
     & ~ & $\hat{N}_{CRC}^*$ & 500.3 & 23.3 & - & - & -   \\ \cline{1-8} 

    \end{tabular*}
    \begin{tablenotes}
        \item[$^{a}$] \footnotesize $Se_1, Sp_1=0.9$, $Se_2, Sp_2=0.95$, $N_{true}=N_{tot}\times p$
        \item[$^{b}$] \footnotesize $\hat{N}_{CRC}$ shows results calculated based on closed-form estimator in equation (\ref{eq_6}) and $\hat{N}_{CRC}^*$ refers to the numerical MLE
        \item[$^{c}$] The Wald-based CI for $\hat{N}_{RS}$ is evaluated by multiplying equation (\ref{eq_3}) by $N_{tot}^2$. The Wald-based CI for $\hat{N}_{CRC}$ is determined using equation (\ref{eq_8}), along with a proposed FPC-adjusted Bayesian credible interval \textbf{(Bold)}
  \end{tablenotes}
\end{table}

Table \ref{table_2} summarizes the simulation results with $N_{tot}$=1,000 and low misclassification level (e.g., $Se_1,Sp_1=0.9$; $Se_2,Sp_2=0.95$). {In this and other simulation settings, the estimation results based on Stream 1 only were heavily biased due to non-representative sampling (results not shown). W}e compare the CRC estimators with the random sampling-based estimator $\hat{N}_{RS}$ justified by the corresponding pre-specified sensitivity and specificity parameters. For the CRC estimators, we report the results for the numerical MLE $\hat{N}_{CRC}^*$ for $N$ along with the closed-form estimator $\hat{N}_{CRC}$ based on equation (\ref{eq_6}). As mentioned previously, a standard error to accompany the numerical $\hat{N}_{CRC}^*$ is not directly available and thus we only report the average point estimate as well as its empirical standard deviation (SD) in the table. While the numerical estimator $\hat{N}_{CRC}^*$ provides better precision in some cases, the difference is very slight and the closed-form estimator $\hat{N}_{CRC}$ is much more convenient for use in practice.

The simulation results in Table \ref{table_2} indicate that all three estimators are virtually unbiased, while the CRC estimators show a clear improvement in estimation precision. {In particular, the improvements in SD range from approximately 10\% of the mean (SD reduced from 53.8 to 44.6 with a mean of 100 in Table \ref{table_2}, top) to 0.9\% of the mean (SD reduced from 27.9 to 23.4 with a mean of 500 in Table \ref{table_2}, bottom). These gains in SD relative to the mean tend to be greatest with realistically small Stream 2 sampling rates. In addition, one can also assess efficiency in terms of improvements in interval width. Reductions in mean width achieved via the proposed FPC-adjusted Bayesian credible interval are typically close to 20\% or more (relative to intervals based on the random sample alone) across most scenarios in Table \ref{table_2}.} The credible interval also provides better coverage compared to the Wald-type confidence interval in most settings especially when the {sample size ($n_2$)} is limited. {While these two intervals behaved similarly when the sample size of the anchor stream increased to 300 (i.e., a 30\% sampling rate), the proposed credible interval is favored overall for its reliable coverage under small anchor stream sample sizes (e.g., 50-100). }

Comparing the low misclassification setting in Table \ref{table_2} with the medium and high settings in Table \ref{table_3} and Table \ref{table_4} respectively, it is clear that as the misclassification level increases, the estimated standard errors and the widths of the intervals become larger. The point estimate of $\hat{N}_{CRC}$ {and $\hat{N}_{RS}$} exhibit minor upward bias for the low prevalence and sampling rate scenario ($p=0.1$, $n_2=50$) at high misclassification level (Table \ref{table_4}) due to the thresholding of any negative prevalence estimates to zero. {It is worth noting that $\hat{N}_{CRC}$ shows a comparatively lesser extent of bias when compared to $\hat{N}_{RS}$.} In the meantime, the proposed Bayesian credible interval approach still provides reliable intervals (i.e., nearly nominal coverage) to accompany the disease case count estimation. {Its width advantages are apparent relative to the interval based on the random sample alone in all cases, and also in most cases relative to the Wald-type CRC interval.}

\begin{table}
    \centering
    \caption{Comparing the Performance of Estimators with $N_{tot}$=1,000 and Medium Misclassification Level $^a$  }   \label{table_3}
    \begin{tabular*}{5in}{@{\extracolsep{\fill}}cccccccc@{\extracolsep{\fill}}}
    \hline
    Prevalence  & Sample & \multirow{2}{*}{Estimator $^{b}$} & \multirow{2}{*}{Mean} & \multirow{2}{*}{SD} & \multirow{2}{*}{Avg. SE} & \multirow{2}{*}{Avg. width $^{c}$} & CI Coverage\\
    $p$ & Size $(n_2)$ & & & & & &  (\%)\\ \hline

    & \multirow{3}{*}{50}  & $\hat{N}_{RS}$ & 101.3 & 62.9 & 66.8 & 261.9 & 95.1   \\ 
     & ~ & $\hat{N}_{CRC}$ & 99.9 & 54.3 & 56.7 & 222.3 \textbf{(206.0)} & 92.7 \textbf{(95.0)}   \\ 
    ~ & ~ & $\hat{N}_{CRC}^*$ & 103.5 & 49.2 & - & - & -   \\ \cline{2-8}
    
     & \multirow{3}{*}{100} & $\hat{N}_{RS}$ & 100.2 & 47.2 & 46.9 & 184.0 & 94.7   \\ 
    0.1  & ~ & $\hat{N}_{CRC}$ & 101.5 & 38.2 & 40.4 & 158.4 \textbf{(142.2)} & 97.4 \textbf{(94.4)}  \\ 
     & ~ & $\hat{N}_{CRC}^*$ & 101.4 & 38.3  & - & - & -   \\ \cline{2-8}
    
    & \multirow{3}{*}{300}  & $\hat{N}_{RS}$ & 100.0 & 26.1 & 26.0 & 102.1 & 95.3   \\ 
     & ~ & $\hat{N}_{CRC}$ & 100.2 & 23.2 & 23.5 & 92.3 \textbf{(89.9)} & 95.2 \textbf{(95.0)}  \\ 
    & ~ & $\hat{N}_{CRC}^*$ & 100.2 & 23.1 & - & - & -   \\  \cline{1-8}

    & \multirow{3}{*}{50}  & $\hat{N}_{RS}$ & 300.3 & 82.1 & 82.4 & 323.1 & 93.0    \\ 
     & ~ & $\hat{N}_{CRC}$ & 300.8 & 68.3 & 67.9 & 266.0 \textbf{(256.1)} & 93.4 \textbf{(95.0)}   \\ 
    ~ & ~ & $\hat{N}_{CRC}^*$ & 300.7 & 68.3  & - & - & -   \\ \cline{2-8}
    
    & \multirow{3}{*}{100} & $\hat{N}_{RS}$ & 300.0 & 57.9 & 57.4 & 225.0 & 94.8  \\ 
    0.3  & ~ & $\hat{N}_{CRC}$ &   299.5 & 48.2 & 47.7 & 187.0 \textbf{(184.5)} & 94.0 \textbf{(94.8)}  \\ 
    ~ & ~ & $\hat{N}_{CRC}^*$ & 299.5 & 48.1 & - & - & -   \\ \cline{2-8}
    
    & \multirow{3}{*}{300}  & $\hat{N}_{RS}$ & 300.0 & 31.2 & 31.0 & 121.4 & 95.0  \\ 
     & ~ & $\hat{N}_{CRC}$ & 299.9 & 26.8 & 27.1 & 106.2 \textbf{(106.4)} & 94.8 \textbf{(95.2)}   \\ 
     & ~ & $\hat{N}_{CRC}^*$ &  299.8 & 26.7 & - & - & -   \\ \cline{1-8}

     & \multirow{3}{*}{50}  & $\hat{N}_{RS}$ & 500.9 & 86.7 & 87.0 & 341.0 & 94.3   \\ 
     & ~ & $\hat{N}_{CRC}$ & 498.6 & 72.5 & 72.1 & 282.5 \textbf{(273.3)} & 94.3 \textbf{(95.1)}   \\ 
    ~ & ~ & $\hat{N}_{CRC}^*$ &  498.6 & 72.5   & - & - & -   \\ \cline{2-8}     
    
    & \multirow{3}{*}{100} & $\hat{N}_{RS}$ & 500.2 & 61.7 & 60.5 & 237.1 & 94.7  \\ 
    0.5  & ~ & $\hat{N}_{CRC}$ &  501.3 & 51.3 & 50.5 & 198.1 \textbf{(194.9)} & 94.6 \textbf{(94.9)}  \\ 
    ~ & ~ & $\hat{N}_{CRC}^*$ & 501.3 & 51.3 & - & - & -   \\ \cline{2-8}
    
    & \multirow{3}{*}{300}  & $\hat{N}_{RS}$ & 500.4 & 32.6 & 32.4 & 127.2 & 95.2  \\ 
     & ~ & $\hat{N}_{CRC}$ &  499.9 & 27.6 & 28.4 & 111.3 \textbf{(111.6)} & 95.5 \textbf{(95.7)}  \\ 
     & ~ & $\hat{N}_{CRC}^*$ & 499.9 & 27.5 & - & - & -   \\ \cline{1-8}

    \end{tabular*}
    \begin{tablenotes}
        \item[$^{a}$] \footnotesize $Se_1, Sp_1=0.85$, $Se_2, Sp_2=0.9$, $N_{true}=N_{tot}\times p$
        \item[$^{b}$] \footnotesize $\hat{N}_{CRC}$ shows results calculated based on closed-form estimator in equation (\ref{eq_6}) and $\hat{N}_{CRC}^*$ refers to the numerical MLE
        \item[$^{c}$] The Wald-based CI for $\hat{N}_{RS}$ is evaluated by multiplying equation (\ref{eq_3}) by $N_{tot}^2$. The Wald-based CI for $\hat{N}_{CRC}$ is determined using equation (\ref{eq_8}), along with a proposed FPC-adjusted Bayesian credible interval \textbf{(Bold)}
  \end{tablenotes}
\end{table}

\begin{table}
    \centering
    \caption{Comparing the Performance of Estimators with $N_{tot}$=1,000 and High Misclassification Level $^a$  }
    \label{table_4}
    \begin{tabular*}{5in}{@{\extracolsep{\fill}}cccccccc@{\extracolsep{\fill}}}
    \hline
    Prevalence  & Sample & \multirow{2}{*}{Estimator $^{b}$} & \multirow{2}{*}{Mean} & \multirow{2}{*}{SD} & \multirow{2}{*}{Avg. SE} & \multirow{2}{*}{Avg. width $^{c}$} & CI Coverage\\
    $p$ & Size $(n_2)$ & & & & & &  (\%)\\ \hline

    & \multirow{3}{*}{50}  & $\hat{N}_{RS}$ & 103.8 & 76.0 & 82.7 & 324.2 & 97.9   \\ 
     & ~ & $\hat{N}_{CRC}$ & 101.6 & 68.1 & 71.8 & 281.6 \textbf{(238.3)} & 96.7 \textbf{(93.6)}   \\ 
    ~ & ~ & $\hat{N}_{CRC}^*$ & 107.9 & 60.5 & - & - & -   \\ \cline{2-8}
    
     & \multirow{3}{*}{100} & $\hat{N}_{RS}$ & 100.3 & 56.5 & 58.3 & 228.5 & 95.1   \\ 
     0.1 & ~ & $\hat{N}_{CRC}$ & 99.0 & 50.0 & 51.2 & 200.7 \textbf{(183.2)} & 94.3 \textbf{(94.3)}  \\ 
     & ~ & $\hat{N}_{CRC}^*$ & 103.6 & 46.1  & - & - & -   \\ \cline{2-8}
    
    & \multirow{3}{*}{300}  & $\hat{N}_{RS}$ & 99.9 & 33.4 & 32.8 & 128.6 & 94.5    \\ 
     & ~ & $\hat{N}_{CRC}$ & 100.1 & 29.6 & 30.1 & 118.0 \textbf{(110.0)} & 95.4 \textbf{(94.5)}  \\ 
    & ~ & $\hat{N}_{CRC}^*$ & 100.0 & 29.5 & - & - & -   \\  \cline{1-8}
       
     & \multirow{3}{*}{50}  & $\hat{N}_{RS}$ & 298.6 & 95.8 & 95.8 & 375.4 & 92.4    \\ 
     & ~ & $\hat{N}_{CRC}$ & 298.7 & 81.2 & 80.1 & 314.2 \textbf{(305.7)} & 93.5 \textbf{(94.1)}   \\ 
    ~ & ~ & $\hat{N}_{CRC}^*$ & 299.1 & 80.3  & - & - & -   \\ \cline{2-8}
    
    & \multirow{3}{*}{100} & $\hat{N}_{RS}$ & 302.2 & 67.1 & 67.1 & 262.9 & 94.9  \\ 
     0.3  & ~ & $\hat{N}_{CRC}$ &   301.1 & 56.5 & 56.8 & 222.6 \textbf{(217.8)} & 95.1 \textbf{(95.3)}  \\ 
    ~ & ~ & $\hat{N}_{CRC}^*$ & 301.1 & 56.5 & - & - & -   \\ \cline{2-8}
    
   & \multirow{3}{*}{300}  & $\hat{N}_{RS}$ & 300.6 & 37.0 & 36.9 & 144.5 & 94.4  \\ 
     & ~ & $\hat{N}_{CRC}$ &  300.7 & 32.6 & 32.8 & 128.5 \textbf{(128.4)} & 94.9 \textbf{(95.0)}   \\ 
     & ~ & $\hat{N}_{CRC}^*$ &  300.7 & 32.4 & - & - & -   \\ \cline{1-8}

    & \multirow{3}{*}{50}  & $\hat{N}_{RS}$ & 499.3 & 100.5 & 99.8 & 391.1 & 93.4  \\ 
     & ~ & $\hat{N}_{CRC}$ &  499.8 & 82.8 & 83.2 & 326.0 \textbf{(314.5)} & 94.2 \textbf{(95.4)}   \\ 
    ~ & ~ & $\hat{N}_{CRC}^*$ &  499.9 & 82.8   & - & - & -   \\ \cline{2-8}
    
    & \multirow{3}{*}{100} & $\hat{N}_{RS}$ & 500.0 & 70.0 & 69.7 & 273.1 & 94.9  \\ 
    0.5  & ~ & $\hat{N}_{CRC}$ &  499.3 & 59.2 & 58.6 & 229.7 \textbf{(225.9)} & 94.4 \textbf{(94.9)}   \\ 
    ~ & ~ & $\hat{N}_{CRC}^*$ & 499.4 & 59.1 & - & - & -   \\ \cline{2-8}
    
    & \multirow{3}{*}{300} & $\hat{N}_{RS}$ & 500.0 & 38.3 & 38.1 & 149.3 & 94.6   \\ 
     & ~ & $\hat{N}_{CRC}$ &  500.2 & 33.2 & 33.6 & 131.9 \textbf{(131.9)} & 95.2 \textbf{(95.2)}  \\ 
     & ~ & $\hat{N}_{CRC}^*$ & 500.3 & 32.9 & - & - & -   \\ \cline{1-8}
    
    \end{tabular*}
    \begin{tablenotes}
        \item[$^{a}$] \footnotesize $Se_1, Sp_1=0.8$, $Se_2, Sp_2=0.85$, $N_{true}=N_{tot}\times p$
        \item[$^{b}$] \footnotesize $\hat{N}_{CRC}$ shows results calculated based on closed-form estimator in equation (\ref{eq_6}) and $\hat{N}_{CRC}^*$ refers to the numerical MLE
        \item[$^{c}$] The Wald-based CI for $\hat{N}_{RS}$ is evaluated by multiplying equation (\ref{eq_3}) by $N_{tot}^2$. The Wald-based CI for $\hat{N}_{CRC}$ is determined using equation (\ref{eq_8}), along with a proposed FPC-adjusted Bayesian credible interval \textbf{(Bold)}
  \end{tablenotes}
\end{table}

{With respect to Tables \ref{table_2}-\ref{table_4}, we note that} the improvement in estimation precision and reduced interval widths were achieved on the basis of the anchor stream design, {with modest (e.g., 50-100)} anchor stream samples collected from a target population {of size 1,000}. A more expanded set of simulation scenarios for the population sizes $N_{tot}=$(200, {10,000}) can be found in Appendix B of the {Supplementary Material} \citep{Ge2024}, {leading to similar conclusions}.

\section{Numerical Example}\label{section_4}

{The design and estimation approaches introduced in Section \ref{section_2} show advantages conceptually and empirically through extensive simulation studies. However, implementing these new tools practically for population-level disease monitoring in accessible closed populations requires strict adherence to the anchor stream design to ensure that diagnostic data from Stream 2 are obtained independently of Stream 1. Given that this design remains newly advocated \citep{Lyles2022a,Lyles2022b, Ge2023}, we provide here an example based in part on simulated and in part on real data. }

Specifically, for the illustration and reproducibility of the proposed approach for case count estimation, we present a numerical example using simulated COVID-19 surveillance data under the anchor stream design. This example is based on one of our simulation settings with parameters $N_{tot}=1,000$, disease prevalence $p=0.1$ and diseased case count $N=100$. The symptom proportion and symptom-specific disease prevalences remain consistent with the values used in the simulation study.

{While synthetic data are utilized to generate individual testing and outcome data, our example is grounded in real validation data to assess the performance of commonly used test kits.} In this illustration, Stream 1 is assumed to arise from voluntary-based testing via rapid antigen test assays, while Stream 2 (the anchor stream) arises from conducting RT-qPCR tests \citep{Adams2020} upon a relatively small random sample of population members ($n_2=100$). To define the performance of the testing tools, we adopt validation data (see Table \ref{table_5}) from \cite{Murakami2023} and \cite{Casati2022} for Stream 1 and Stream 2, respectively. Thus, the assumed sensitivity and specificity for Stream 1 are $Se_1=63\%$, $Sp_1=99.8\%$; in contrast, these parameters for Stream 2 are $Se_2=94\%$, $Sp_2=100\%$. 

\begin{table}
\caption{Validation Data for Test Performance in Two Data Streams\label{table_5}}%
\begin{tabular*}{\columnwidth}{@{\extracolsep\fill}lcccc@{\extracolsep\fill}}
\hline
& & \multicolumn{3}{@{}c@{}}{True Disease Status} \\
\cline{3-5}%
&& + & \textminus & Total  \\ 
\hline
Rapid Antigen   & +   & 65 (63\%)  & 1 (0.2\%) & 66 \\
Test (for     & \textminus  & 38 (37\%)  & 552 (99.8\%) & 590 \\
Stream 1)$^{a}$     & Total   & 103 (100\%)  & 553 (100\%) & 656  \\
\hline
RT- qPCR   & +   & 89 (94\%)  & 0 (0\%) & 89 \\
Test (for    & \textminus  & 6 (6\%)  & 100 (100\%) & 106 \\
Stream 2)$^{b}$     & Total   & 95 (100\%)  & 100 (100\%) & 195  \\
\hline
\end{tabular*}
\begin{tablenotes}%
\item[$^{a}$] True validation data is available from \cite{Murakami2023}.
\item[$^{b}$] True validation data is available from \cite{Casati2022}.
\end{tablenotes}
\end{table}

The primary example data was generated under the prescribed anchor stream design conditions, in the sense that the random sample of size 100 (Stream 2) was collected agnostically with respect to the voluntary-based samples in Stream 1 among a population of size 1,000. Hence, we obtained for illustration a single set of simulated observed cell counts (see Table \ref{table_1}), as follows: $n_1=3$, $n_2=12$, $n_3=0$, $n_4=2$, $n_5=27$, $n_6=130$, $n_7=6$, $n_8=77$, $n_9=743$.

Given the observed cell counts and setting the misclassification parameters equal to the estimates based on Table \ref{table_5}, we can directly calculate the case count estimates and corresponding confidence (or credible) intervals using the approaches introduced in Section \ref{section_2}. With access to validation data for estimating the misclassification parameters as in Table \ref{table_5}, however, we recommend {the approach introduced in Section \ref{section_2.5} that} adapts the multiple imputation (MI) paradigm \citep{Rubin1987} to account for uncertainty in these parameters. 

{The results of this illustration are displayed in Table \ref{table_6}.} As expected, the results obtained via the MI paradigm demonstrate higher uncertainty compared to the directly implemented proposed method. In practice, if the misclassification parameters are deemed reliable for inclusion in the analysis, we recommend implementing the approach outlined {in the Section \ref{section_2.2}-\ref{section_2.4}} directly. Alternatively, if validation data are available from the manufacturer or collected as part of the surveillance effort, we recommend applying the MI paradigm {introduced in Section \ref{section_2.5}} to account for uncertainty in the $Se$ and $Sp$ parameters. 

To fully utilize this example, including R code to facilitate case count estimation and data simulation, we direct readers to Appendix C of the Supplementary Material \citep{Ge2024} or the following GitHub link: https://github.com/lge-biostat/CRC-with-SeSp.

\begin{table}
\caption{Numerical Example: Case Count Estimates with $N_{tot}=1,000$ and $N_{true}=100$ \label{table_6}}%
\begin{tabular*}{\columnwidth}{@{\extracolsep\fill}lcccc@{\extracolsep\fill}}
\hline
Estimator & Mean & SE & 95\% CI & Width  \\ 
\hline
$\hat{N}_{RS}$$^{a}$    & 117.4   & 32.0  & (54.8, 180.1) & 125.3 \\
$\hat{N}_{CRC}$$^{b}$  & 111.5  & 24.7  & (63.2, 159.9), \textbf{(75.1, 172.6)}  & 96.7, \textbf{97.5} \\
$\hat{N}_{RS, MI}$$^{c}$    & 113.7   & 33.3  & (48.4, 178.9) & 130.6 \\
$\hat{N}_{CRC, MI}$$^{d}$  & 108.2  & 26.0  & (57.2, 159.2), \textbf{(68.5, 172.7)} & 102.1, \textbf{104.2} \\
\hline
\end{tabular*}
\begin{tablenotes}%
\item[$^{a}$] The Wald-based CI for $\hat{N}_{RS}$ is evaluated by multiplying eqn.(\ref{eq_3}) by ${N}_{tot}^2$.
\item[$^{b}$] The Wald-based CI for $\hat{N}_{CRC}$ is determined using eqn.(\ref{eq_8}), along with a proposed FPC-adjusted Bayesian credible interval \textbf{(Bold)}.
\item[$^{c}$] The Wald-based CI for $\hat{N}_{RS, MI}$ is evaluated by multiplying eqn.(\ref{eq_3}) by ${N}_{tot}^2$ through $M=100$ multiple imputations.
\item[$^{d}$] The Wald-based CI for $\hat{N}_{CRC,MI}$ is determined using eqn.(\ref{eq_8}) through $M=100$ multiple imputations, along with a proposed FPC-adjusted Bayesian credible interval \textbf{(Bold)} based on $S=1,000$ Dirichlet posterior draws for each of the $M$ imputations.
\end{tablenotes}
\end{table}

\section{Discussion}

In this article, we propose a more flexible capture-recapture strategy for accelerating infectious disease monitoring, accounting for imperfect diagnostic or test results. We believe that this work is timely and well-motivated for monitoring the prevalence or case counts of infectious diseases such as COVID-19 or measles among a registered population, e.g. schools, communities, and geographic regions, when a diagnostic device or test-kit leverages an imperfect test for rapid results. To adjust for misclassified diagnostic signals, we extend recently proposed anchor stream design and methods \citep{Lyles2022a,Lyles2022b} for CRC analysis in epidemiological disease surveillance without misclassification by incorporating pre-specified sensitivity and specificity information from manufactured test kits.

When focusing on disease monitoring in a closed and registered population from which a representative random sample can be drawn and misclassification parameters associated with the diagnostic device or test-kit are available from the manufacturer, the proposed method for anchor stream-based CRC analysis is straightforward to implement in practice. It is important to note, however, that the anchor stream sample must be drawn carefully to assure not only its representativeness but also its independence relative to the voluntary testing stream \citep{Lyles2022a, Lyles2022b}. Our empirical studies indicate that leveraging a relatively small anchor stream sample together with arbitrarily non-representative voluntary test results can unlock a much more precise estimator of the true case count or prevalence in the target population than could be achieved through either sample alone. Along with existing disease surveillance data streams, this method can provide accurate and timely results. {While we acknowledge likely difficulties in obtaining complete test results on the anchor stream sample in attempts to scale this design to large populations (e.g., census tracts or counties), it is intriguing to see that the benefits are there to be harnessed if the implementation challenges could be overcome.}

During the COVID-19 pandemic or in other infectious disease monitoring efforts, the proposed CRC strategy may be useful for application among registered populations for periodic monitoring of infectious disease prevalence in an efficient and economical way. The key is to have reliable information about the misclassification parameters (sensitivity and specificity) for each surveillance effort. In developing the proposed methods, we generally assumed that the $Se$ and $Sp$ parameters were accurately specified by test-kit manufacturers for utilization in the analysis. Nevertheless, unknown misclassification parameters present a common challenge in practice. As demonstrated in our numerical example, one can straightforwardly extend the proposed approach in the event that validation data are available to account for the fact that $Se$ and $Sp$ parameters are unknown or are inaccurate due to practical issues with implementation of the diagnostic device or test-kit. Leveraging the extra information from internal or external validation data will lead to additional uncertainty in the estimation, but also adds a layer of robustness and expands the practical uses of this CRC strategy in solving real-world problems. Further investigation of validation study designs in this disease surveillance context could be a useful topic for future research.

\begin{funding}
This work was supported by the National Institute of Health (NIH)/National Institute of Allergy and Infectious Diseases (P30AI050409; Del Rio PI), the NIH/National Center for Advancing Translational Sciences (UL1TR002378; Taylor PI), the NIH/National Cancer Institute  (R01CA234538; Ward/Lash MPIs), and the NIH/National Cancer Institute (R01CA266574; Lyles/Waller MPIs).\vspace*{-8pt}
\end{funding}



\begin{supplement}
\stitle{Appendices}
\sdescription{This document presents the following contents: Appendix A. Details of extra terms involved in the FPC-adjusted variance estimator; Appendix B. Additional Simulation Results; Appendix C. Code File for this paper.}
\end{supplement}

\bibliographystyle{imsart-nameyear} 
\bibliography{references.bib}       




\end{document}


\begin{frontmatter}
\title{Supplement to ``Utilizing a Capture-Recapture Strategy to Accelerate Infectious Disease Surveillance”}

\begin{aug}
\author[A,B]{\fnms{Lin}~\snm{Ge}\ead[label=e1]{lge$\_$biostat@hotmail.com}\orcid{0000-0002-8180-9615}},
\author[A]{\fnms{Yuzi}~\snm{Zhang}\ead[label=e2]{yuzi.zhang@emory.edu}\orcid{0000-0003-1805-6910}}
\author[A]{\fnms{Lance}~\snm{Waller}\ead[label=e3]{lwaller@emory.edu}\orcid{0000-0001-5002-8886}}
\and
\author[A]{\fnms{Robert}~\snm{Lyles}\ead[label=e4]{rlyles@emory.edu}}
\address[A]{Department of Biostatistics and Bioinformatics, Emory University\printead[presep={,\ }]{e2,e3,e4}}

\address[B]{Department of Biostatistics and Epidemiology, Harvard University\printead[presep={,\ }]{e1}}

\end{aug}

\end{frontmatter}



\begin{appendix}

\section{Details of extra terms involved in the FPC-adjusted variance estimator}\label{Apendix_1}

The FPC-adjusted variance estimator is derived following the same strategy that yields the variance of the more standard bias-corrected prevalence estimator under finite population sampling in equation (3) (Ge et al., 2023). Each extra term comprising the final variance estimator has the following form:  
\begin{align}
    \hat{V}_2(\hat{\pi}_{ij}) = FPC_{ij}\hat{V}_1(\hat{\pi}_{ij}) + \hat{V}_{extra}^{ij}, ~~ i,j=0,1 
\end{align}
Here, $\hat{V}_{extra}^{ij}$ is derived as follows:
\begin{align}
    \hat{V}_{extra}^{11} &= \frac{1}{(Se_2+Sp_2-1)^2}\frac{1}{N_{1}}[\hat{\pi}_{11} Se_2(1-Se_2)+  (1-\hat{\pi}_{11})Sp_2(1-Sp_2)] \\
    \hat{V}_{extra}^{10} &= \frac{1}{(Se_1+Sp_1-1)^2}\frac{1}{N_{1}}[\hat{\pi}_{10} Se_1(1-Se_1)+  (1-\hat{\pi}_{10})Sp_1(1-Sp_1)] \\
    \hat{V}_{extra}^{01} &= \frac{1}{(Se_2+Sp_2-1)^2}\frac{1}{N_{tot}-N_{1}}[\hat{\pi}_{01} Se_2(1-Se_2)+  (1-\hat{\pi}_{01})Sp_2(1-Sp_2)] 
\end{align}

\newpage

\section{Additional Simulation Results}

\begin{table}[ht]
    \centering
    \caption{Comparing the Performance of Estimators with $N_{tot}$=200 and Low Misclassification Level $^a$ }
    \label{table_s1}
    \begin{tabular*}{5in}{@{\extracolsep{\fill}}cccccccc@{\extracolsep{\fill}}}
    \hline
    Prevalence  & Sample & \multirow{2}{*}{Estimator $^{b}$} & \multirow{2}{*}{Mean} & \multirow{2}{*}{SD} & \multirow{2}{*}{Avg. SE} & \multirow{2}{*}{Avg. width $^{c}$} & CI Coverage\\
    $p$ & Size $(n_2)$ & & & & & &  (\%)\\ \hline
    
~ & ~ & $\hat{N}_{RS}$ & 22.3 & 20.9 & 21.8 & 85.6 & 78.2\\
~ & 10 & $\hat{N}_{CRC}$ & 21.2 & 18.8 & 17.8 & 69.9\textbf{ (72.2)} & 75.9\textbf{ (97.2)}\\
~ & ~ & $\hat{N}_{CRC}^*$ & 23.5 & 16.6 & - & - & -\\
\cline{2-8}
~ & ~ & $\hat{N}_{RS}$ & 20.3 & 15.6 & 16.1 & 63.3 & 94.3\\
0.1 & 20 & $\hat{N}_{CRC}$ & 20.3 & 13.6 & 13.3 & 52.0\textbf{ (51.4)} & 84.4\textbf{ (97.3)}\\
~ & ~ & $\hat{N}_{CRC}^*$ & 21.3 & 12.3 & - & - & -\\
\cline{2-8}
~ & ~ & $\hat{N}_{RS}$ & 20.0 & 9.0 & 9.0 & 35.1 & 94.3\\
~ & 60 & $\hat{N}_{CRC}$ & 20.0 & 7.7 & 7.9 & 30.9\textbf{ (30.8)} & 94.3\textbf{ (94.9)}\\
~ & ~ & $\hat{N}_{CRC}^*$ & 20.2 & 7.4 & - & - & -\\
\cline{1-8}
~ & ~ & $\hat{N}_{RS}$ & 60.0 & 32.0 & 31.6 & 123.9 & 86.8\\
~ & 10 & $\hat{N}_{CRC}$ & 60.5 & 26.3 & 24.2 & 94.8\textbf{ (89.7)} & 85.4\textbf{ (95.7)}\\
~ & ~ & $\hat{N}_{CRC}^*$ & 61.3 & 25.3 & - & - & -\\
\cline{2-8}
~ & ~ & $\hat{N}_{RS}$ & 59.9 & 22.2 & 22.2 & 87.0 & 90.7\\
0.3 & 20 & $\hat{N}_{CRC}$ & 60.3 & 18.2 & 18.0 & 70.6\textbf{ (67.5)} & 91.1\textbf{ (94.5)}\\
~ & ~ & $\hat{N}_{CRC}^*$ & 60.4 & 18.0 & - & - & -\\
\cline{2-8}
~ & ~ & $\hat{N}_{RS}$ & 60.1 & 11.9 & 11.7 & 46.0 & 93.8\\
~ & 60 & $\hat{N}_{CRC}$ & 60.2 & 10.0 & 10.0 & 39.4\textbf{ (39.1)} & 94.6\textbf{ (95.6)}\\
~ & ~ & $\hat{N}_{CRC}^*$ & 60.2 & 9.9 & - & - & -\\
\cline{1-8}
~ & ~ & $\hat{N}_{RS}$ & 100.1 & 34.3 & 34.4 & 134.8 & 90.1\\
~ & 10 & $\hat{N}_{CRC}$ & 99.7 & 29.0 & 26.9 & 105.4\textbf{ (94.5)} & 88.3\textbf{ (94.3)}\\
~ & ~ & $\hat{N}_{CRC}^*$ & 99.9 & 28.6 & - & - & -\\
\cline{2-8}
~ & ~ & $\hat{N}_{RS}$ & 99.5 & 23.9 & 23.9 & 93.5 & 90.4\\
0.5 & 20 & $\hat{N}_{CRC}$ & 99.8 & 20.0 & 19.8 & 77.5\textbf{ (72.1)} & 92.2\textbf{ (94.9)}\\
~ & ~ & $\hat{N}_{CRC}^*$ & 99.8 & 20.0 & - & - & -\\
\cline{2-8}
~ & ~ & $\hat{N}_{RS}$ & 99.9 & 12.5 & 12.5 & 49.0 & 94.9\\
~ & 60 & $\hat{N}_{CRC}$ & 100.0 & 10.6 & 10.8 & 42.5\textbf{ (41.9)} & 94.8\textbf{ (95.4)}\\
~ & ~ & $\hat{N}_{CRC}^*$ & 100.0 & 10.6 & - & - & - \\ \hline
    \end{tabular*}
    \begin{tablenotes}
        \item[$^{a}$] \footnotesize $Se_1, Sp_1=0.9$, $Se_2, Sp_2=0.95$, $N_{true}=N_{tot}\times p$
        \item[$^{b}$] \footnotesize $\hat{N}_{CRC}$ shows results calculated based on closed-form estimator in equation (6) and $\hat{N}_{CRC}^*$ refers to the numerical MLE
        \item[$^{c}$] The Wald-based CI for $\hat{N}_{RS}$ is evaluated by multiplying equation (3) by $N_{tot}^2$. The Wald-based CI for $\hat{N}_{CRC}$ is determined using equation (8), along with a proposed FPC-adjusted Bayesian credible interval \textbf{(Bold)}
  \end{tablenotes}
\end{table}

\begin{table}
    \centering
    \caption{Comparing the Performance of Estimators with $N_{tot}$=200 and Medium Misclassification Level $^a$  }   \label{table_s2}
    \begin{tabular*}{5in}{@{\extracolsep{\fill}}cccccccc@{\extracolsep{\fill}}}
    \hline
    Prevalence  & Sample & \multirow{2}{*}{Estimator $^{b}$} & \multirow{2}{*}{Mean} & \multirow{2}{*}{SD} & \multirow{2}{*}{Avg. SE} & \multirow{2}{*}{Avg. width $^{c}$} & CI Coverage\\
    $p$ & Size $(n_2)$ & & & & & &  (\%)\\ \hline
    
 ~ & ~ & $\hat{N}_{RS}$ & 23.8 & 26.5 & 28.3 & 111.0 & 86.0\\
~ & 10 & $\hat{N}_{CRC}$ & 23.5 & 23.2 & 23.6 & 92.7\textbf{ (81.7)} & 82.9\textbf{ (97.2)}\\
~ & ~ & $\hat{N}_{CRC}^*$ & 25.9 & 21.1 & - & - & -\\
\cline{2-8}
~ & ~ & $\hat{N}_{RS}$ & 21.6 & 18.9 & 20.7 & 81.0 & 96.6\\
0.1 & 20 & $\hat{N}_{CRC}$ & 21.1 & 16.8 & 17.8 & 69.6\textbf{ (60.1)} & 92.3\textbf{ (95.4)}\\
~ & ~ & $\hat{N}_{CRC}^*$ & 23.0 & 14.8 & - & - & -\\
\cline{2-8}
~ & ~ & $\hat{N}_{RS}$ & 20.2 & 11.4 & 11.6 & 45.6 & 93.2\\
~ & 60 & $\hat{N}_{CRC}$ & 20.3 & 10.3 & 10.5 & 41.3\textbf{ (38.3)} & 94.6\textbf{ (94.2)}\\
~ & ~ & $\hat{N}_{CRC}^*$ & 20.8 & 9.5 & - & - & -\\
\cline{1-8}
~ & ~ & $\hat{N}_{RS}$ & 59.6 & 36.1 & 36.4 & 142.6 & 88.0\\
~ & 10 & $\hat{N}_{CRC}$ & 60.3 & 30.5 & 28.4 & 111.3\textbf{ (103.6)} & 87.8\textbf{ (94.8)}\\
~ & ~ & $\hat{N}_{CRC}^*$ & 61.8 & 28.4 & - & - & -\\
\cline{2-8}
~ & ~ & $\hat{N}_{RS}$ & 60.4 & 25.2 & 25.7 & 100.7 & 94.1\\
0.3 & 20 & $\hat{N}_{CRC}$ & 60.5 & 21.3 & 21.4 & 83.7\textbf{ (79.5)} & 92.6\textbf{ (95.1)}\\
~ & ~ & $\hat{N}_{CRC}^*$ & 60.8 & 20.7 & - & - & -\\
\cline{2-8}
~ & ~ & $\hat{N}_{RS}$ & 59.7 & 13.9 & 13.8 & 54.3 & 94.5\\
~ & 60 & $\hat{N}_{CRC}$ & 59.8 & 11.8 & 12.1 & 47.5\textbf{ (47.1)} & 94.6\textbf{ (95.6)}\\
~ & ~ & $\hat{N}_{CRC}^*$ & 59.8 & 11.7 & - & - & -\\
\cline{1-8}
~ & ~ & $\hat{N}_{RS}$ & 100.5 & 38.7 & 38.8 & 152.2 & 90.0\\
~ & 10 & $\hat{N}_{CRC}$ & 100.3 & 32.7 & 30.6 & 120.0\textbf{ (107.5)} & 88.9\textbf{ (94.5)}\\
~ & ~ & $\hat{N}_{CRC}^*$ & 100.7 & 31.7 & - & - & -\\
\cline{2-8}
~ & ~ & $\hat{N}_{RS}$ & 99.6 & 27.0 & 27.1 & 106.1 & 89.7\\
0.5 & 20 & $\hat{N}_{CRC}$ & 99.8 & 22.8 & 22.7 & 88.8\textbf{ (82.8)} & 92.4\textbf{ (95.1)}\\
~ & ~ & $\hat{N}_{CRC}^*$ & 99.9 & 22.6 & - & - & -\\
\cline{2-8}
~ & ~ & $\hat{N}_{RS}$ & 100.2 & 14.5 & 14.5 & 56.9 & 94.0\\
~ & 60 & $\hat{N}_{CRC}$ & 100.1 & 12.5 & 12.7 & 49.9\textbf{ (49.2)} & 95.0\textbf{ (95.7)}\\
~ & ~ & $\hat{N}_{CRC}^*$ & 100.2 & 12.4 & - & - & -   \\ \hline
    \end{tabular*}
    \begin{tablenotes}
        \item[$^{a}$] \footnotesize $Se_1, Sp_1=0.85$, $Se_2, Sp_2=0.9$, $N_{true}=N_{tot}\times p$
        \item[$^{b}$] \footnotesize $\hat{N}_{CRC}$ shows results calculated based on closed-form estimator in equation (6) and $\hat{N}_{CRC}^*$ refers to the numerical MLE
        \item[$^{c}$] The Wald-based CI for $\hat{N}_{RS}$ is evaluated by multiplying equation (3) by $N_{tot}^2$. The Wald-based CI for $\hat{N}_{CRC}$ is determined using equation (8), along with a proposed FPC-adjusted Bayesian credible interval \textbf{(Bold)}
  \end{tablenotes}
\end{table}

\begin{table}
    \centering
    \caption{Comparing the Performance of Estimators with $N_{tot}$=200 and High Misclassification Level $^a$  }
    \label{table_s3}
    \begin{tabular*}{5in}{@{\extracolsep{\fill}}cccccccc@{\extracolsep{\fill}}}
    \hline
    Prevalence  & Sample & \multirow{2}{*}{Estimator $^{b}$} & \multirow{2}{*}{Mean} & \multirow{2}{*}{SD} & \multirow{2}{*}{Avg. SE} & \multirow{2}{*}{Avg. width $^{c}$} & CI Coverage\\
    $p$ & Size $(n_2)$ & & & & & &  (\%)\\ \hline
    
   ~ & ~ & $\hat{N}_{RS}$ & 26.3 & 28.7 & 35.7 & 139.8 & 91.0\\
~ & 10 & $\hat{N}_{CRC}$ & 24.2 & 26.5 & 29.9 & 117.2\textbf{ (91.1)} & 99.0\textbf{ (97.9)}\\
~ & ~ & $\hat{N}_{CRC}^*$ & 27.5 & 24.0 & - & - & -\\
\cline{2-8}
~ & ~ & $\hat{N}_{RS}$ & 22.3 & 22.1 & 25.7 & 100.8 & 97.8\\
0.1 & 20 & $\hat{N}_{CRC}$ & 22.0 & 20.1 & 22.6 & 88.5\textbf{ (68.2)} & 95.3\textbf{ (94.6)}\\
~ & ~ & $\hat{N}_{CRC}^*$ & 24.4 & 17.9 & - & - & -\\
\cline{2-8}
~ & ~ & $\hat{N}_{RS}$ & 20.4 & 13.8 & 14.7 & 57.4 & 97.5\\
~ & 60 & $\hat{N}_{CRC}$ & 20.1 & 12.7 & 13.5 & 52.8\textbf{ (44.8)} & 98.0\textbf{ (94.0)}\\
~ & ~ & $\hat{N}_{CRC}^*$ & 21.1 & 11.4 & - & - & -\\
\cline{1-8}
~ & ~ & $\hat{N}_{RS}$ & 61.7 & 40.3 & 42.6 & 167.0 & 90.2\\
~ & 10 & $\hat{N}_{CRC}$ & 61.2 & 35.6 & 34.0 & 133.5\textbf{ (117.9)} & 89.3\textbf{ (93.5)}\\
~ & ~ & $\hat{N}_{CRC}^*$ & 63.6 & 32.4 & - & - & -\\
\cline{2-8}
~ & ~ & $\hat{N}_{RS}$ & 60.0 & 29.5 & 29.9 & 117.3 & 94.5\\
0.3 & 20 & $\hat{N}_{CRC}$ & 60.3 & 25.6 & 25.3 & 99.1\textbf{ (92.1)} & 91.7\textbf{ (93.8)}\\
~ & ~ & $\hat{N}_{CRC}^*$ & 61.2 & 24.1 & - & - & -\\
\cline{2-8}
~ & ~ & $\hat{N}_{RS}$ & 60.1 & 16.6 & 16.5 & 64.6 & 93.8\\
~ & 60 & $\hat{N}_{CRC}$ & 59.9 & 14.7 & 14.7 & 57.5\textbf{ (56.7)} & 94.4\textbf{ (94.9)}\\
~ & ~ & $\hat{N}_{CRC}^*$ & 60.0 & 14.5 & - & - & -\\
\cline{1-8}
~ & ~ & $\hat{N}_{RS}$ & 100.1 & 44.5 & 44.5 & 174.4 & 88.8\\
~ & 10 & $\hat{N}_{CRC}$ & 100.1 & 37.4 & 35.4 & 138.6\textbf{ (122.7)} & 88.9\textbf{ (94.3)}\\
~ & ~ & $\hat{N}_{CRC}^*$ & 100.9 & 35.6 & - & - & -\\
\cline{2-8}
~ & ~ & $\hat{N}_{RS}$ & 99.4 & 31.1 & 31.2 & 122.2 & 96.5\\
0.5 & 20 & $\hat{N}_{CRC}$ & 99.7 & 26.4 & 26.3 & 103.1\textbf{ (96.1)} & 92.8\textbf{ (94.4)}\\
~ & ~ & $\hat{N}_{CRC}^*$ & 99.8 & 26.0 & - & - & -\\
\cline{2-8}
~ & ~ & $\hat{N}_{RS}$ & 99.8 & 17.1 & 17.1 & 66.8 & 93.4\\
~ & 60 & $\hat{N}_{CRC}$ & 99.8 & 14.9 & 15.1 & 59.1\textbf{ (58.2)} & 94.5\textbf{ (95.3)}\\
~ & ~ & $\hat{N}_{CRC}^*$ & 99.8 & 14.8 & - & - & -   \\ \hline
    \end{tabular*}
    \begin{tablenotes}
        \item[$^{a}$] \footnotesize $Se_1, Sp_1=0.8$, $Se_2, Sp_2=0.85$, $N_{true}=N_{tot}\times p$
        \item[$^{b}$] \footnotesize $\hat{N}_{CRC}$ shows results calculated based on closed-form estimator in equation (6) and $\hat{N}_{CRC}^*$ refers to the numerical MLE
        \item[$^{c}$] The Wald-based CI for $\hat{N}_{RS}$ is evaluated by multiplying equation (3) by $N_{tot}^2$. The Wald-based CI for $\hat{N}_{CRC}$ is determined using equation (8), along with a proposed FPC-adjusted Bayesian credible interval \textbf{(Bold)}
  \end{tablenotes}
\end{table}


\begin{table}
    \centering
    \caption{Comparing the Performance of Estimators with $N_{tot}$=10,000 and Low Misclassification Level $^a$ }
    \label{table_s4}
    \begin{tabular*}{5in}{@{\extracolsep{\fill}}cccccccc@{\extracolsep{\fill}}}
    \hline
    Prevalence  & Sample & \multirow{2}{*}{Estimator $^{b}$} & \multirow{2}{*}{Mean} & \multirow{2}{*}{SD} & \multirow{2}{*}{Avg. SE} & \multirow{2}{*}{Avg. width $^{c}$} & CI Coverage\\
    $p$ & Size $(n_2)$ & & & & & &  (\%)\\ \hline
    
~ & ~ & $\hat{N}_{RS}$ & 997.1 & 540.7 & 536.8 & 2104.3 & 92.4\\
~ & 50 & $\hat{N}_{CRC}$ & 994.4 & 442.6 & 428.5 & 1679.8\textbf{ (1679.6)} & 91.4\textbf{ (93.4)}\\
~ & ~ & $\hat{N}_{CRC}^*$ & 1155.7 & 469.4 & - & - & -\\
\cline{2-8}
~ & ~ & $\hat{N}_{RS}$ & 1003.3 & 376.9 & 382.7 & 1500.2 & 94.4\\
0.1 & 100 & $\hat{N}_{CRC}$ & 1001.9 & 308.6 & 310.2 & 1216.2\textbf{ (1212.7)} & 93.6\textbf{ (94.3)}\\
~ & ~ & $\hat{N}_{CRC}^*$ & 1004.2 & 303.1 & - & - & -\\
\cline{2-8}
~ & ~ & $\hat{N}_{RS}$ & 1004.2 & 216.3 & 220.5 & 864.2 & 95.7\\
~ & 300 & $\hat{N}_{CRC}$ & 997.7 & 178.0 & 179.7 & 704.6\textbf{ (701.7)} & 94.3\textbf{ (95.1)}\\
~ & ~ & $\hat{N}_{CRC}^*$ & 997.9 & 177.7 & - & - & -\\
\cline{1-8}
~ & ~ & $\hat{N}_{RS}$ & 2985.6 & 735.1 & 729.5 & 2859.5 & 93.0\\
~ & 50 & $\hat{N}_{CRC}$ & 2988.7 & 585.9 & 576.7 & 2260.7\textbf{ (2206.8)} & 93.4\textbf{ (95.2)}\\
~ & ~ & $\hat{N}_{CRC}^*$ & 3030.9 & 524.1 & - & - & -\\
\cline{2-8}
~ & ~ & $\hat{N}_{RS}$ & 2999.8 & 522.3 & 515.8 & 2022.0 & 94.3\\
0.3 & 100 & $\hat{N}_{CRC}$ & 3003.6 & 417.6 & 412.9 & 1618.5\textbf{ (1594.9)} & 94.0\textbf{ (94.8)}\\
~ & ~ & $\hat{N}_{CRC}^*$ & 3004.5 & 417.1 & - & - & -\\
\cline{2-8}
~ & ~ & $\hat{N}_{RS}$ & 2991.7 & 293.8 & 295.5 & 1158.3 & 95.0\\
~ & 300 & $\hat{N}_{CRC}$ & 2993.9 & 237.2 & 237.4 & 930.7\textbf{ (924.7)} & 94.7\textbf{ (94.9)}\\
~ & ~ & $\hat{N}_{CRC}^*$ & 2993.8 & 237.2 & - & - & -\\
\cline{1-8}
~ & ~ & $\hat{N}_{RS}$ & 5020.7 & 790.5 & 783.9 & 3072.9 & 93.6\\
~ & 50 & $\hat{N}_{CRC}$ & 5016.0 & 647.7 & 635.2 & 2489.9\textbf{ (2404.5)} & 93.9\textbf{ (94.8)}\\
~ & ~ & $\hat{N}_{CRC}^*$ & 4995.6 & 682.1 & - & - & -\\
\cline{2-8}
~ & ~ & $\hat{N}_{RS}$ & 5006.7 & 551.5 & 553.3 & 2169.1 & 94.7\\
0.5 & 100 & $\hat{N}_{CRC}$ & 5001.6 & 448.7 & 452.2 & 1772.7\textbf{ (1737.9)} & 94.8\textbf{ (95.4)}\\
~ & ~ & $\hat{N}_{CRC}^*$ & 5002.5 & 448.0 & - & - & -\\
\cline{2-8}
~ & ~ & $\hat{N}_{RS}$ & 4995.6 & 315.2 & 316.9 & 1242.0 & 94.6\\
~ & 300 & $\hat{N}_{CRC}$ & 4997.3 & 258.3 & 259.7 & 1017.9\textbf{ (1009.4)} & 95.1\textbf{ (95.3)}\\
~ & ~ & $\hat{N}_{CRC}^*$ & 4997.6 & 258.3 & - & - & -   \\ \hline
    \end{tabular*}
    \begin{tablenotes}
        \item[$^{a}$] \footnotesize $Se_1, Sp_1=0.9$, $Se_2, Sp_2=0.95$, $N_{true}=N_{tot}\times p$
        \item[$^{b}$] \footnotesize $\hat{N}_{CRC}$ shows results calculated based on closed-form estimator in equation (6) and $\hat{N}_{CRC}^*$ refers to the numerical MLE
        \item[$^{c}$] The Wald-based CI for $\hat{N}_{RS}$ is evaluated by multiplying equation (3) by $N_{tot}^2$. The Wald-based CI for $\hat{N}_{CRC}$ is determined using equation (8), along with a proposed FPC-adjusted Bayesian credible interval \textbf{(Bold)}
  \end{tablenotes}
\end{table}

\begin{table}
    \centering
    \caption{Comparing the Performance of Estimators with $N_{tot}$=10,000 and Medium Misclassification Level $^a$  }   \label{table_s5}
    \begin{tabular*}{5in}{@{\extracolsep{\fill}}cccccccc@{\extracolsep{\fill}}}
    \hline
    Prevalence  & Sample & \multirow{2}{*}{Estimator $^{b}$} & \multirow{2}{*}{Mean} & \multirow{2}{*}{SD} & \multirow{2}{*}{Avg. SE} & \multirow{2}{*}{Avg. width $^{c}$} & CI Coverage\\
    $p$ & Size $(n_2)$ & & & & & &  (\%)\\ \hline
    
~ & ~ & $\hat{N}_{RS}$ & 1017.5 & 645.1 & 673.8 & 2641.3 & 95.4\\
~ & 50 & $\hat{N}_{CRC}$ & 1013.9 & 562.1 & 562.0 & 2202.9\textbf{ (2048.9)} & 92.5\textbf{ (93.9)}\\
~ & ~ & $\hat{N}_{CRC}^*$ & 1260.0 & 495.5 & - & - & -\\
\cline{2-8}
~ & ~ & $\hat{N}_{RS}$ & 997.8 & 472.8 & 477.2 & 1870.6 & 94.1\\
0.1 & 100 & $\hat{N}_{CRC}$ & 999.4 & 407.9 & 402.1 & 1576.2\textbf{ (1529.4)} & 93.6\textbf{ (93.8)}\\
~ & ~ & $\hat{N}_{CRC}^*$ & 1012.9 & 385.2 & - & - & -\\
\cline{2-8}
~ & ~ & $\hat{N}_{RS}$ & 998.3 & 278.1 & 275.2 & 1078.9 & 94.3\\
~ & 300 & $\hat{N}_{CRC}$ & 1000.5 & 238.5 & 233.4 & 915.0\textbf{ (909.1)} & 93.9\textbf{ (94.3)}\\
~ & ~ & $\hat{N}_{CRC}^*$ & 1001.1 & 237.2 & - & - & -\\
\cline{1-8}
~ & ~ & $\hat{N}_{RS}$ & 2999.4 & 845.4 & 834.8 & 3272.6 & 92.0\\
~ & 50 & $\hat{N}_{CRC}$ & 3006.2 & 690.3 & 674.1 & 2642.5\textbf{ (2573.1)} & 92.8\textbf{ (94.4)}\\
~ & ~ & $\hat{N}_{CRC}^*$ & 3102.4 & 473.1 & - & - & -\\
\cline{2-8}
~ & ~ & $\hat{N}_{RS}$ & 2995.4 & 590.8 & 589.9 & 2312.2 & 94.3\\
0.3 & 100 & $\hat{N}_{CRC}$ & 3002.1 & 483.6 & 481.7 & 1888.4\textbf{ (1857.5)} & 94.4\textbf{ (94.9)}\\
~ & ~ & $\hat{N}_{CRC}^*$ & 3002.6 & 482.2 & - & - & -\\
\cline{2-8}
~ & ~ & $\hat{N}_{RS}$ & 2999.7 & 342.4 & 338.7 & 1327.8 & 94.8\\
~ & 300 & $\hat{N}_{CRC}$ & 3000.7 & 277.5 & 277.8 & 1089.0\textbf{ (1080.4)} & 94.3\textbf{ (94.8)}\\
~ & ~ & $\hat{N}_{CRC}^*$ & 3000.7 & 277.6 & - & - & -\\
\cline{1-8}
~ & ~ & $\hat{N}_{RS}$ & 4979.2 & 873.1 & 882.6 & 3459.8 & 93.8\\
~ & 50 & $\hat{N}_{CRC}$ & 4985.7 & 724.2 & 717.2 & 2811.5\textbf{ (2714.7)} & 93.7\textbf{ (94.7)}\\
~ & ~ & $\hat{N}_{CRC}^*$ & 4863.4 & 887.9 & - & - & -\\
\cline{2-8}
~ & ~ & $\hat{N}_{RS}$ & 4983.8 & 621.8 & 623.0 & 2442.2 & 94.7\\
0.5 & 100 & $\hat{N}_{CRC}$ & 4990.3 & 507.3 & 510.3 & 2000.5\textbf{ (1961.4)} & 95.0\textbf{ (95.2)}\\
~ & ~ & $\hat{N}_{CRC}^*$ & 4989.8 & 508.0 & - & - & -\\
\cline{2-8}
~ & ~ & $\hat{N}_{RS}$ & 4993.5 & 357.8 & 357.4 & 1400.9 & 95.0\\
~ & 300 & $\hat{N}_{CRC}$ & 4999.8 & 292.0 & 294.4 & 1154.1\textbf{ (1143.8)} & 95.2\textbf{ (95.2)}\\
~ & ~ & $\hat{N}_{CRC}^*$ & 5000.0 & 291.9 & - & - & -   \\ \hline
    \end{tabular*}
    \begin{tablenotes}
        \item[$^{a}$] \footnotesize $Se_1, Sp_1=0.85$, $Se_2, Sp_2=0.9$, $N_{true}=N_{tot}\times p$
        \item[$^{b}$] \footnotesize $\hat{N}_{CRC}$ shows results calculated based on closed-form estimator in equation (6) and $\hat{N}_{CRC}^*$ refers to the numerical MLE
        \item[$^{c}$] The Wald-based CI for $\hat{N}_{RS}$ is evaluated by multiplying equation (3) by $N_{tot}^2$. The Wald-based CI for $\hat{N}_{CRC}$ is determined using equation (8), along with a proposed FPC-adjusted Bayesian credible interval \textbf{(Bold)}
  \end{tablenotes}
\end{table}

\begin{table}
    \centering
    \caption{Comparing the Performance of Estimators with $N_{tot}$=10,000 and High Misclassification Level $^a$  }
    \label{table_s6}
    \begin{tabular*}{5in}{@{\extracolsep{\fill}}cccccccc@{\extracolsep{\fill}}}
    \hline
    Prevalence  & Sample & \multirow{2}{*}{Estimator $^{b}$} & \multirow{2}{*}{Mean} & \multirow{2}{*}{SD} & \multirow{2}{*}{Avg. SE} & \multirow{2}{*}{Avg. width $^{c}$} & CI Coverage\\
    $p$ & Size $(n_2)$ & & & & & &  (\%)\\ \hline
    
~ & ~ & $\hat{N}_{RS}$ & 1040.8 & 771.9 & 831.7 & 3260.3 & 97.9\\
~ & 50 & $\hat{N}_{CRC}$ & 1022.0 & 689.3 & 709.1 & 2779.7\textbf{ (2372.4)} & 95.6\textbf{ (93.2)}\\
~ & ~ & $\hat{N}_{CRC}^*$ & 1465.2 & 388.9 & - & - & -\\
\cline{2-8}
~ & ~ & $\hat{N}_{RS}$ & 1018.6 & 576.9 & 590.1 & 2313.4 & 95.4\\
0.1 & 100 & $\hat{N}_{CRC}$ & 1015.8 & 508.6 & 509.2 & 1996.0\textbf{ (1833.7)} & 94.0\textbf{ (93.1)}\\
~ & ~ & $\hat{N}_{CRC}^*$ & 1044.1 & 465.3 & - & - & -\\
\cline{2-8}
~ & ~ & $\hat{N}_{RS}$ & 997.1 & 339.3 & 340.0 & 1332.8 & 95.0\\
~ & 300 & $\hat{N}_{CRC}$ & 1000.4 & 292.3 & 294.8 & 1155.5\textbf{ (1140.8)} & 94.9\textbf{ (93.7)}\\
~ & ~ & $\hat{N}_{CRC}^*$ & 1003.0 & 286.5 & - & - & -\\
\cline{1-8}
~ & ~ & $\hat{N}_{RS}$ & 3008.3 & 964.7 & 968.1 & 3795.1 & 92.3\\
~ & 50 & $\hat{N}_{CRC}$ & 3012.6 & 813.9 & 795.4 & 3117.8\textbf{ (3028.1)} & 93.5\textbf{ (94.7)}\\
~ & ~ & $\hat{N}_{CRC}^*$ & 3062.1 & 722.1 & - & - & -\\
\cline{2-8}
~ & ~ & $\hat{N}_{RS}$ & 3000.9 & 694.2 & 683.8 & 2680.5 & 93.6\\
0.3 & 100 & $\hat{N}_{CRC}$ & 2997.5 & 582.4 & 566.4 & 2220.1\textbf{ (2181.0)} & 93.8\textbf{ (94.1)}\\
~ & ~ & $\hat{N}_{CRC}^*$ & 2996.2 & 565.1 & - & - & -\\
\cline{2-8}
~ & ~ & $\hat{N}_{RS}$ & 3002.4 & 399.5 & 393.2 & 1541.4 & 94.3\\
~ & 300 & $\hat{N}_{CRC}$ & 3005.1 & 332.7 & 327.9 & 1285.2\textbf{ (1275.5)} & 94.3\textbf{ (94.6)}\\
~ & ~ & $\hat{N}_{CRC}^*$ & 3005.1 & 332.7 & - & - & -\\
\cline{1-8}
~ & ~ & $\hat{N}_{RS}$ & 5003.8 & 1018.4 & 1008.6 & 3953.8 & 93.5\\
~ & 50 & $\hat{N}_{CRC}$ & 5000.2 & 847.8 & 821.8 & 3221.6\textbf{ (3112.7)} & 93.5\textbf{ (94.5)}\\
~ & ~ & $\hat{N}_{CRC}^*$ & 4808.5 & 1159.4 & - & - & -\\
\cline{2-8}
~ & ~ & $\hat{N}_{RS}$ & 5015.5 & 724.9 & 712.4 & 2792.6 & 94.3\\
0.5 & 100 & $\hat{N}_{CRC}$ & 5009.4 & 592.7 & 586.8 & 2300.2\textbf{ (2254.2)} & 94.2\textbf{ (94.5)}\\
~ & ~ & $\hat{N}_{CRC}^*$ & 5006.5 & 598.1 & - & - & -\\
\cline{2-8}
~ & ~ & $\hat{N}_{RS}$ & 4999.1 & 405.9 & 409.4 & 1604.7 & 94.6\\
~ & 300 & $\hat{N}_{CRC}$ & 4997.0 & 340.1 & 338.9 & 1328.3\textbf{ (1316.4)} & 94.7\textbf{ (94.9)}\\
~ & ~ & $\hat{N}_{CRC}^*$ & 4997.1 & 340.1 & - & - & -  \\ \hline
    \end{tabular*}
    \begin{tablenotes}
        \item[$^{a}$] \footnotesize $Se_1, Sp_1=0.8$, $Se_2, Sp_2=0.85$, $N_{true}=N_{tot}\times p$
        \item[$^{b}$] \footnotesize $\hat{N}_{CRC}$ shows results calculated based on closed-form estimator in equation (6) and $\hat{N}_{CRC}^*$ refers to the numerical MLE
        \item[$^{c}$] The Wald-based CI for $\hat{N}_{RS}$ is evaluated by multiplying equation (3) by $N_{tot}^2$. The Wald-based CI for $\hat{N}_{CRC}$ is determined using equation (8), along with a proposed FPC-adjusted Bayesian credible interval \textbf{(Bold)}
  \end{tablenotes}
\end{table}
\end{appendix}